\documentclass[twoside,twocolumn]{article}

\usepackage[sc]{mathpazo} 
\usepackage[T1]{fontenc} 
\usepackage{microtype} 

\usepackage[english]{babel} 

\usepackage[hmarginratio=1:1,top=32mm,columnsep=20pt]{geometry} 
\usepackage[hang, small,labelfont=bf,up,textfont=it,up]{caption} 
\usepackage{booktabs} 

\usepackage{enumitem} 
\setlist[itemize]{noitemsep} 

\usepackage{abstract} 

\usepackage{titlesec} 
\renewcommand\thesection{\Roman{section}} 
\renewcommand\thesubsection{\roman{subsection}} 
\titleformat{\section}[block]{\large\scshape\centering}{\thesection.}{1em}{} 
\titleformat{\subsection}[block]{\large}{\thesubsection.}{1em}{} 

\usepackage{fancyhdr} 
\usepackage{titling} 

\usepackage{amsmath}
\usepackage{amsfonts}
\usepackage{amssymb}
\usepackage{graphicx}

\newcommand{\vect}[1]{\boldsymbol{#1}} 

\pretitle{\begin{center}\Huge\bfseries} 
\posttitle{\end{center}} 
\title{Random Wave Function Collapse} 
\author{%
\textsc{I. Mayergoyz}\\
\normalsize ECE Department, University of Maryland, College Park, MD 20742  
}
\date{} 
\renewcommand{\maketitlehookd}{%
\begin{abstract}
\noindent Instability-induced random branching of deterministic dynamics is discussed as a possible mechanism of random wave function collapse. In the case of two level systems, the Born probability rule emerges as the simplest linear solution to the equation for measurement probabilities. 
\end{abstract}
}

\begin{document}

\maketitle


It is well-known that determinism and randomness coexist in the mathematical structure of quantum mechanics. This coexistence leads to the paradox of random wave function collapse. The random wave function collapse follows from the Measurement Postulate of quantum mechanics which is often called the \textbf{Born rule}. According to this postulate, during  measurements, wave functions collapse into  specific eigenfunctions corresponding to  measured eigenvalues of Hermitian operators. Since results of measurements are probabilistic in nature, this implies the random collapse of original wave functions. This random wave function collapse is seemingly inconsistent with the purely deterministic time-evolution of wave functions described by the time-dependent Schr\"odinger equation. This inconsistency of the Born rule with the time-dependent Schr\"odinger equation has been the source of lingering controversy and ongoing scientific debate. 

It is worthwhile to mention the remarkable \textbf{universality} of the Born rule. Indeed, the probability of wave function collapse is determined by the same rule regardless of the physical nature of microscopic systems, their states and the physical quantities being measured. This probability (often called the Born probability) is equal to $|c_n|^2$ where $c_n$ is a coefficient in the expansion of a state wave function with respect to the joint (common) eigenfunctions $\varphi_n(\mathbf{x})$ of Hermitian operators corresponding to a complete set of the simultaneously measurable physical quantities. 

The paradox of random wave function collapse resulted in the emergence of different interpretations of quantum mechanics. Examples of such interpretations include ``hidden variables" theories, the ``many worlds" interpretation, ``consistent histories" or ``decoherent histories" theories, and ``ensemble" models. It was also suggested to introduce random and nonlinear terms into the time-dependent Schr\"odinger equation to account for the random wave function collapse during measurements. The critical discussion of these and other theories along with appropriate references can be found in [1,2].

The prevailing point of view is that the quantum mechanical nature of the measurement process and the mechanism of random wave function collapse have not been clearly understood yet. It would be desirable to demonstrate that the Born rule is derivable. However, it is hardly doable. The mathematical reason is that a macroscopic measuring device has a huge number of microscopic degrees of freedom. This makes it impossible to solve the time-dependent Schr\"odinger equation and to trace the emergence of the Born rule for various physical measurements. 

It is apparent that during measurements appreciable amplification (enhancement) of microscopic effects must occur in order to make them tangible to a macroscopic observer. It is natural to assume that this amplification is due to certain instabilities involved in the measurement process. It turns out that under some circumstances, instabilities may lead to randomness in deterministic dynamics. It is along this line of reasoning that the measurement process is analyzed in the following discussion. This discussion of quantum mechanical measurements does not involve any change in the foundation of quantum mechanics.

Consider a microscopic system described by the wave function $\tilde{\psi}(\mathbf{x},t)$ where $\mathbf{x}$ stands for all microscopic degrees of freedom. To be specific, we assume that energy along with some other physical quantities are being simultaneously measured. For this reason, we represent $\tilde{\psi}(\mathbf{x},0)$ as follows:
\begin{equation}
\tilde{\psi}(\mathbf{x},0) = \sum_n c_n \varphi_n(\mathbf{x}),
\label{eqA.1}
\end{equation}
where $\varphi_n(\mathbf{x})$ are joint eigenfunctions of the Hamiltonian and other operators of simultaneously measurable physical quantities. It is apparent that
\begin{equation}
\int \left| \tilde{\psi} (\mathbf{x}, t) \right|^2 \, d\mathbf{x} = \sum_n |c_n|^2 =1.
\label{eqA.2}
\end{equation}
Next, we shall use the notation $\psi (\mathbf{q}, \mathbf{x},t)$ for the wave function of the microscopic system and measuring device. Here $\mathbf{q}$ stands for all microscopic degrees of freedom of the macroscopic measuring device. It is also apparent that 
\begin{equation}
\int \left| \psi (\mathbf{q}, \mathbf{x},t) \right|^2 \, d\mathbf{q} d\mathbf{x} =1.
\label{eqA.3} 
\end{equation}
Suppose that the measurement is initiated at time $t=0$. Since the macroscopic device and microscopic system do not interact before $t=0$, the initial wave function of the overall system can be represented as the following product
\begin{equation}
\psi (\mathbf{q},\mathbf{x},0) = A(\mathbf{q}) \tilde{\psi}(\mathbf{x},0).
\label{eqA.4}
\end{equation}
It is apparent from formulas (\ref{eqA.2}), (\ref{eqA.3}) and (\ref{eqA.4}) that
\begin{equation}
\int |A(\mathbf{q})|^2 \, d\mathbf{q}=1.
\label{eqA.5}
\end{equation}
We shall use the following expansion for the overall wave function 
\begin{equation}
\psi (\mathbf{q}, \mathbf{x},t) = \sum_n a_n (\mathbf{q},t) \varphi_n (\mathbf{x}) e^{-i \frac{\mathcal{E}_n}{\hbar}t}.
\label{eqA.6}
\end{equation}
From the last formula, we find that
\begin{align}
\begin{split}
&\left| \psi (\mathbf{q}, \mathbf{x},t) \right|^2 \\ = \sum_n \sum_m a_n^* (\mathbf{q},t) &a_m(\mathbf{q},t) \varphi_n^* (\mathbf{x}) \varphi_m(\mathbf{x}) e^{i \frac{\mathcal{E}_n - \mathcal{E}_m}{\hbar}t}.
\end{split}
\label{eqA.7}
\end{align}
By taking into account the orthonormality of functions $\varphi_n(\mathbf{x})$, we derive
\begin{equation}
\int \left| \psi (\mathbf{q}, \mathbf{x}, t) \right|^2 \, d\mathbf{x} = \sum_n |a_n (\mathbf{q},t) |^2.
\label{eqA.8}
\end{equation}
Now, from formulas (\ref{eqA.3}) and (\ref{eqA.8}) we conclude that
\begin{equation}
\sum_n \int |a_n (\mathbf{q},t)|^2 \, d\mathbf{q}=1.
\label{eqA.9}
\end{equation}
Next, we introduce the following functions
\begin{equation}
b_n(t) = \left[ \int | a_n (\mathbf{q},t
)|^2 \,d\mathbf{q} \right]^\frac{1}{2}.
\label{eqA.10}
\end{equation}
It is clear from (\ref{eqA.9}) that
\begin{equation}
\sum_n b_n^2(t)=1.
\label{eqA.11}
\end{equation}
Formulas (\ref{eqA.10}) and (\ref{eqA.11}) imply that the dynamics of the overall wave function $\tilde{\psi}(\mathbf{q}, \mathbf{x},t)$ during any measurement process can be mapped into the dynamics of  functions $b_n(t)$ on the ``positive" part $S_+$ of the unit sphere, that is, the part of the unit sphere with positive coordinates.

Next, by using formulas (\ref{eqA.1}), (\ref{eqA.4}) and (\ref{eqA.6}) we find that
\begin{align}
\begin{split}
\psi(\mathbf{q}, \mathbf{x},0) &= \sum_n a_n (\mathbf{q},0) \varphi_n (\mathbf{x}) 
\\ &= A(\mathbf{q}) \sum_n c_n \varphi_n(\mathbf{x}),
\end{split}
\label{eqA.12}
\end{align}
which leads to
\begin{equation}
a_n (\mathbf{q},0) =A(\mathbf{q}) c_n.
\label{eqA.13}
\end{equation}
From the last formula as well as formulas (\ref{eqA.5}) and (\ref{eqA.10}), we find that
\begin{equation}
b_n(0) = |c_n|,
\label{eqA.14}
\end{equation}
which can be construed as the initial condition for the dynamics of $b_n(t)$ on the unit sphere.

The Born rule and experiments suggest that as a result of measurement the overall wave function $\psi(\mathbf{q},\mathbf{x},t)$ is collapsed with probability
\begin{equation}
P_{n_{0}} = |c_{n_{0}}|^2
\label{eqA.15}
\end{equation}
into the wave function
\begin{equation}
a_{n_0} (\mathbf{q}, \tau) \varphi_{n_0} (\mathbf{x}) e^{-i \frac{\mathcal{E}_{n_0}}{\hbar} \tau},
\label{eqA.16}
\end{equation}
where $\tau$ stands for the time-duration of measurement.

This implies that the deterministic dynamics of $b_n(t)$ on $S_+$ starting from the initial condition (\ref{eqA.14}) randomly collapses into the point of the unit sphere with the coordinates
\begin{equation}
b_{n_0}(\tau) =1, \quad b_n(\tau) =0 \quad \text{for } n \neq n_0.
\label{eqA.17}
\end{equation}
Conversely, it is also apparent that the described random collapse of $b_n$-dynamics on the unit sphere implies the random collapse of the overall wave function $\psi(\mathbf{q}, \mathbf{x},t)$ into the wave function given by the formula (\ref{eqA.16}).

Now, the question can be posed under what conditions the deterministic $b_n$-dynamics on the ``positive" part of the unit sphere randomly collapses into one of the points specified by formula (\ref{eqA.17}). To answer this question, we describe the $b_n$-dynamics on $S_+$ by the equation
\begin{equation}
\frac{d\mathbf{b}(t)}{dt} = \vect{v}(\mathbf{b}(t)),
\label{eqA.18}
\end{equation}
where $\mathbf{b}(t)$ is a vector whose Cartesian components are $b_n(t)$.

Since equation (\ref{eqA.18}) describes the dynamics on the unit sphere, we easily find that
\begin{equation}
\mathbf{b}(t) \cdot \vect{v}(\mathbf{b}(t))=0.
\label{eqA.19}
\end{equation}
Next, consider the case when vector $\vect{v}(\mathbf{b}(t))$ can be decomposed as follows
\begin{equation}
\vect{v} (\mathbf{b}(t)) = - \nabla_{S_+} f(\mathbf{b}(t)) + \mathbf{w}(\mathbf{b}(t)),
\label{eqA.20}
\end{equation}
where
\begin{equation}
\mathbf{w}(\mathbf{b}(t)) \cdot \nabla_{S_+} f(\mathbf{b}(t)) =0.
\label{eqA.21}
\end{equation}
It is assumed in the last two formulas that function $f(\mathbf{b}(t))$ is defined on $S_+$ and $\nabla_{S_+}f(\mathbf{b}(t))$ is the gradient of $f(\mathbf{b}(t))$ along $S_+$.

By using the last two formulas, the equation (\ref{eqA.18}) can be written as
\begin{equation}
\frac{d\mathbf{b}(t)}{dt} = - \nabla_{S_+} f(\mathbf{b}(t)) + \mathbf{w}(\mathbf{b}(t)).
\label{eqA.22a}
\end{equation}

It is also clear that
\begin{equation}
\frac{df(\mathbf{b}(t))}{dt} = \nabla_{S_+}f(\mathbf{b}(t)) \cdot \frac{d \mathbf{b}(t)}{dt}.
\label{eqA.22}
\end{equation}
Now, from the last two formulas, we find
\begin{equation}
\frac{df(\mathbf{b}(t))}{dt} = - \left| \nabla_{S_+} f(\mathbf{b}(t)) \right|^2 <0.
\label{eqA.23}
\end{equation}
This implies that the $\mathbf{b}$-dynamics on $S_+$ results in the monotonic decrease of $f(\mathbf{b}(t))$. This suggests that equation (\ref{eqA.22a}) can be viewed as a generalized Landau-Lifshitz-type equation where the gradient of the $f$-function can be construed as a generalized damping. 

Now, consider the case when $f(\mathbf{b}(t))$ has only one maximum value at the point on $S_+$ specified by the initial conditions (\ref{eqA.14}) and minimum values only at the points specified by formula (\ref{eqA.17}) with all possible values of $n_0$. According to formula (\ref{eqA.23}), points (\ref{eqA.17}) can be viewed as attractors whose basins of attractions may be finely interlaced around the initial point (\ref{eqA.14}). It is assumed here that $f$ is such that there are no attractive (stable) limit cycles and points (\ref{eqA.17}) are the only attractors. Then, it is clear that this $\mathbf{b}$-dynamics is intrinsically unstable. Any tiny perturbation of the initial condition (\ref{eqA.14}) results according to inequality (\ref{eqA.23}) in deterministic  $\mathbf{b}$-dynamics (relaxation) which ends in one of the $f$-minimum points (\ref{eqA.17}). If a tiny perturbation is random, then the random branching of deterministic dynamics occurs and the final destination of the deterministic $\mathbf{b}$-dynamics is random as well.

It is interesting to consider the case of two-level systems. In this case $\mathbf{b}$-dynamics occurs on a ``positive" part $S_+$ of the unit circle (see Figure \ref{fig1} below).
\begin{figure}[ht]
\centerline{\includegraphics[width=2in]{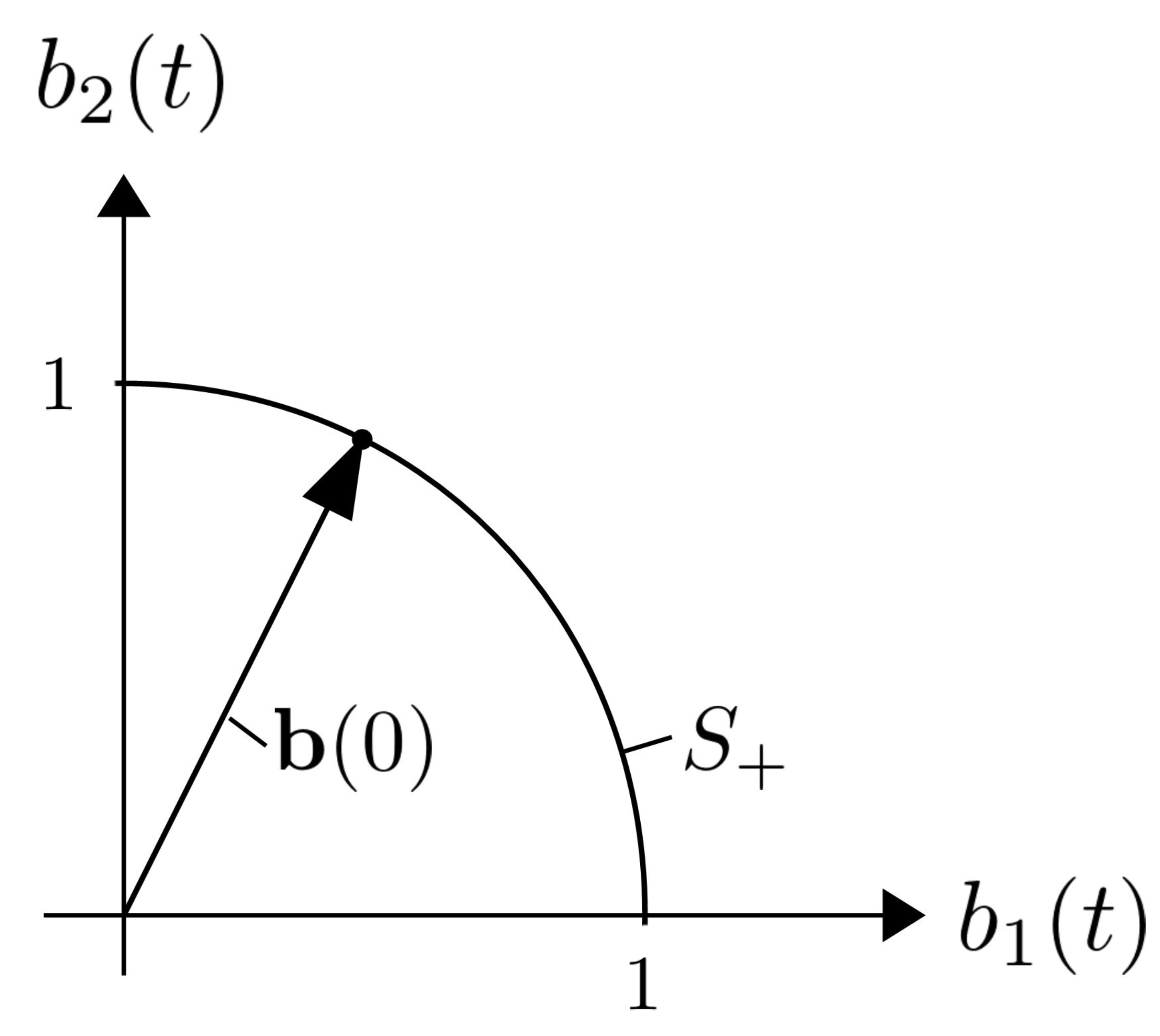}}
\caption{}
\label{fig1}
\end{figure}
Accordingly, the $\mathbf{b}$-dynamics can be represented by the equation 
\begin{equation}
\frac{d\mathbf{b}(t)}{dt} = - \nabla_{S_+}f(\mathbf{b}(t)),
\label{eqA.24}
\end{equation}
where $f(\mathbf{b}(t))$ is a function that has only one maximum value on $S_+$ at the point specified by the initial vector $\mathbf{b}(0)$ and only two minimum values at the points $(1,0)$ and $(0,1)$. By using the same reasoning as before, it is easy to derive from equation (\ref{eqA.24}) that
\begin{equation}
\frac{df(\mathbf{b}(t))}{dt} = - |\nabla_{S_+} f(\mathbf{b}(t))|^2 < 0.
\label{eqA.25}
\end{equation}
Thus, if the deterministic dynamics starts at the point specified by initial vector $\mathbf{b}(0)$, it can only relax to one of the boundary points of $S_+$. This relaxation is deterministic. However, the choice of relaxation trajectory to one of the boundary points is controlled by chance. This leads to the random branching in deterministic $\mathbf{b}$-dynamics and to the randomness in its final destination.

Let $P_1$ and $P_2$ be the probabilities of random collapse to the boundary points $(1,0)$ and $(0,1)$. These probabilities depend on the initial position. Consequently, they depend on $b_1^2(0)$. Thus, we can write
\begin{equation}
P_1(b_1^2(0)) + P_2(b_1^2(0)) =1.
\label{eqA.26}
\end{equation}
Due to symmetry reasons, it can be concluded that
\begin{equation}
P_2(b_1^2(0)) = P_1(1-b_1^2(0)),
\label{eqA.27}
\end{equation}
which leads to
\begin{equation}
P_1(b_1^2(0)) + P_1(1-b_1^2(0))=1.
\label{eqA.28}
\end{equation}
Now, the Born probability rule emerges as the simplest linear solution of equation (\ref{eqA.28}):
\begin{equation}
P_1(b_1^2(0)) = b_1^2(0) = |c_1|^2,
\label{eqA.29}
\end{equation}
which also implies that
\begin{equation}
P_2(b_1^2(0)) = b_2^2(0)=|c_2|^2.
\label{eqA.30}
\end{equation}

There are two possible sources of tiny randomness involved in measurements. One source is due to the fact that there is no way to control (and reproduce exactly) the wave function of a macroscopic measuring device (with a huge number of microscopic degrees of freedom) before the onset of each measurement. For this reason, each measurement is performed with slightly different (and random) initial wave function of the measuring device. However, as it is clear from formula (\ref{eqA.14}), this randomness does not affect the initial conditions for $\mathbf{b}$-dynamics. Another source of randomness is due to the fact that the measuring device and microscopic system (being measured) do not form, strictly speaking, a closed system. In other words, any measurement is performed in the presence of the fluctuating environment and, for this reason, measurements will always be subject to tiny random perturbations from the external environment. These tiny random perturbations may lead to random branching in deterministic $\mathbf{b}$-dynamics and result in the random collapse of the wave function.

It is worthwhile to point out that our analysis has been carried out by using only the unitarity condition for wave function evolution, and no dynamics equation for such evolution has been used. This brings the difficult question of the  mathematical mechanism of instability on the level of the linear dynamic Schr\"odinger equation. It can be conjectured that it may be a parametric instability due to the periodic-in-time and space coefficients in equations for functions $a_n(\mathbf{q},t)$.




\begin{thebibliography}{2} 


\item[{[1]}]Steven Weinberg, Physical Review A, \textbf{85}, 062116 (2012).

\item[{[2]}]Angelo Bassi, Kinjalk Lochan, Seema Satin, Tejinder P. Singh, Hendrick Ulbricht, Review of Modern Physics, vol. 85, 471-527 (2013).
 
\end{thebibliography}
\end{document}